\documentclass[prd,aps,twocolumn,floatfix,preprintnumbers,groupedaddress,nofootinbib]{revtex4-1}
\pdfoutput=1

\newcommand{\smu}[1]{\widetilde{\mu}_{#1}}

\newcommand{\thmu}{\theta_\mu}

\newcommand{\lsim}
{\;\raisebox{-.3em}{$\stackrel{\displaystyle <}{\sim}$}\;}

\usepackage{amsmath}
\usepackage{amssymb, bm}
\usepackage{graphicx}
\usepackage{float}
\usepackage{braket}
\usepackage{hyperref}
\usepackage{pbox}
\usepackage{tabularx}
\usepackage{enumitem}
\usepackage{tikz}\usetikzlibrary{intersections}
\usepackage{epstopdf}
\usepackage[bottom]{footmisc}
\newcommand{\bmat}{\begin{pmatrix}}
\newcommand{\emat}{\end{pmatrix}}
\def\beq{\begin{equation}}
\def\eeq{\end{equation}}
\def\beeq{\begin{eqnarray}}
\def\eeeq{\end{eqnarray}}
\def\bea{\begin{align}}
\def\eea{\end{align}}

\def\gp2{g^{\prime 2}}

\usepackage{color}
\begin{document}   
\preprint{DAMTP-2015-72}
\preprint{IPPP/15/63}
\preprint{DCPT/15/124}
\preprint{TUM-HEP-1027-15}
\title{The ATLAS Di-boson Excess Could Be an $R-$parity Violating Di-smuon Excess} 
\date{\today}
\author{B.~C.~Allanach} 
\email{B.C.Allanach@damtp.cam.ac.uk}
\affiliation{Department of Applied Mathematics and Theoretical Physics, Centre
  for Mathematical Sciences, University of Cambridge, Wilberforce Road,
  Cambridge CB3 0WA, UK} 
\author{P. S. Bhupal Dev} 
\email{bhupal.dev@tum.de}
\affiliation{Physik Department T30d,
Technische Universit\"{a}t M\"{u}nchen, James-Franck-Stra\ss e 1, D-85748 Garching, Germany}
\author{Kazuki Sakurai}
\email{kazuki.sakurai@durham.ac.uk}
\affiliation{Institute for Particle Physics Phenomenology, Ogden Centre for
  Fundamental Physics, Department of Physics, University of Durham, Science
  Laboratories, South Road, Durham DH1 3LE, UK} 

\begin{abstract} 
We propose a new possible explanation of the ATLAS di-boson excess: that it is
due to heavy resonant slepton production, followed by decay into di-smuons. The 
smuon has a mass not too far from the $W$ and $Z$ masses,  and
so it is easily 
confused with $W$ 
or $Z$ bosons after its subsequent decay into di-jets, through a supersymmetry
violating and $R-$parity violating interaction. Such a scenario is not currently
excluded by other constraints and remains to be definitively tested in Run II
of the LHC. 
Such light smuons can easily simultaneously explain 
the discrepancy between the measurement of the anomalous magnetic moment of
the muon and the Standard Model prediction. 
\end{abstract}   
\maketitle 

\section{Introduction}

In 20.3 fb$^{-1}$ of $\sqrt s=8$ TeV LHC data, ATLAS measured an excess with
respect to 
Standard Model (SM) predictions in the
production of di-electroweak gauge bosons $VV$ (where $V=W,Z$) that decay
hadronically~\cite{Aad:2015owa}. The excess 
was at a di-boson invariant mass around 1.8--2 TeV, and occurred in all three decay
channels: $WZ$, $WW$ and $ZZ$ with local significance of 3.4, 2.6 and
2.9$\sigma$, respectively.\footnote{It is interesting to note that a similar
  previous Run I search by CMS also had an excess, albeit milder, around the same mass~\cite{Khachatryan:2014hpa}.} The hadronically decaying di-bosons were identified by using jet mass and sub-jet grooming and mass-drop filtering   techniques~\cite{Butterworth:2008iy}. 
Despite some initial worries about the method of application of such
techniques~\cite{Goncalves:2015yua}, they have so far held up to
re-scrutinization theoretically~\cite{Aad:2015rpa}. ATLAS and CMS analyzed 3.2
and 2.2 fb$^{-1}$ of Run II $\sqrt{s}=13$ TeV data, respectively, and although
no diboson excess above 2$\sigma$ was found, the 
sensitivity was too small to rule out the Run I excess at the 95$\%$
confidence level (CL)~\cite{ATLAS-CONF-2015-073,CMS-PAS-EXO-15-002}. 

There have been many proposals of new physics in order to explain the Run I
excess. 
Most of the early proposals involved the production of various
different types of spin-one resonances~\cite{Fukano:2015hga, Hisano:2015gna,Franzosi:2015zra, Cheung:2015nha,Dobrescu:2015qna,Saavedra:2015rna,Alves:2015mua,Gao:2015irw,Thamm:2015csa,Brehmer:2015cia,Cao:2015lia,
Cacciapaglia:2015eea,Abe:2015jra,Allanach:2015hba,Abe:2015uaa,Carmona:2015xaa,Dobrescu:2015yba,
Chiang:2015lqa,Anchordoqui:2015uea,Bian:2015ota,Kim:2015vba,Lane:2015fza,
Faraggi:2015iaa,Low:2015uha,Arnan:2015csa,Niehoff:2015iaa, Dev:2015pga,Deppisch:2015cua,Bian:2015hda,
Awasthi:2015ota,Li:2015yya,collins:2015wua,Fukano:2015zua}. 
There
were also some attempts involving spin-zero 
~\cite{Cacciapaglia:2015nga, Sanz:2015zha, Chen:2015xql, Omura:2015nwa, Chao:2015wea,  Kim:2015vba, Chiang:2015lqa, Cacciapaglia:2015nga, Fichet:2015yia, Arnan:2015csa, Chen:2015cfa, Sierra:2015zma}, spin-two~\cite{Sanz:2015zha, Kim:2015vba, Fichet:2015yia} as well as composite fermion~\cite{Xue:2015wha} resonances.  
However, none of these proposals involved sparticle resonances from the well-motivated minimal
supersymmetric standard model (MSSM).\footnote{One attempt~\cite{Petersson:2015rza} did use a
  sgoldstino resonance: a spin-zero component of the goldstino. Such a
  scenario requires a fundamental supersymmetry breaking scale at a few TeV.} Here, we wish to
construct a model that 
{\em is}\/ consistent with the MSSM and that explains the ATLAS di-boson
excess, thus potentially additionally solving the technical hierarchy
problem and reinvigorating the hopes of confirming low-scale supersymmetry
(SUSY) in Run II of the LHC. We take advantage of the fact that the ATLAS
di-boson excess only relies on 
the mass of boosted jets in order to identify $W$ and $Z$ bosons. If the
heavy resonance decays instead to other states which have a mass in the
vicinity of 
the $W$ and $Z$ and then each of them decays to di-jets (which, because of the
large 
resonance mass, look like one boosted fat jet with a two sub-jet structure),
this scenario will not be distinguished from the $VV$ resonance in the ATLAS
analysis.   

The rest of the paper is organised as follows: in Section~\ref{sec:prop}, we present the basic idea involving light smuons and test their compatibility with the existing constraints. A general discussion of the smuon masses and mixing is given in Section~\ref{sec:mass}, followed by the mass assignments in our RPV scenario in Section~\ref{sec:ass} and the slepton decay widths in Section~\ref{sec:decay}. A fit to the di-boson excess is presented in Section~\ref{sec:fit}. Some discussions followed by our conclusion are given in Section~\ref{sec:conc}.  

\section{The proposal}\label{sec:prop}

\begin{figure}[b!]
	\centering 
    \includegraphics[width=0.45\textwidth]{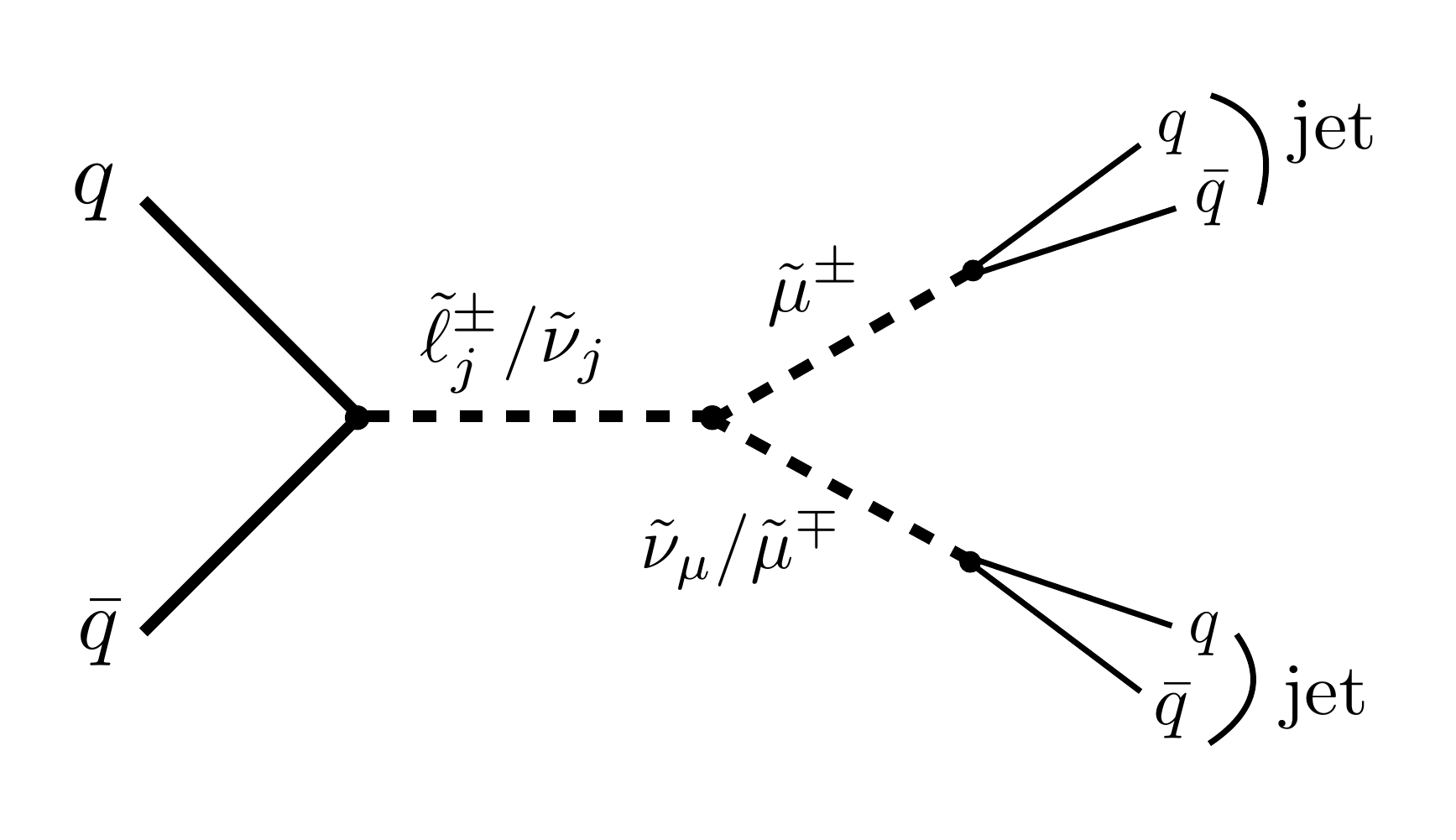}
    \caption{Slepton resonance mimicking the ATLAS di-boson excess. The 
      resonant charged slepton/sneutrino has a mass of around 1.9 TeV and the
      smuons must be in the mass regime 80-105 GeV, such that they would be
      mistaken for $W$ or $Z$'s after their boosted hadronic decay.
\label{fig:main_diag}}
\end{figure}
Our proposal is depicted in Fig.~\ref{fig:main_diag}, where each vertex is $R-$parity violating (RPV). There are three independent
vertices, requiring three different interaction terms in the RPV MSSM. 
We write the relevant part of the RPV superpotential (for a review, see e.g.~\cite{Barbier:2004ez}) 
\beq
W_{\rm LV} \  =  \ \lambda'_{j11} L_j Q_1 \bar D_1 + \lambda'_{2kl} L_2 Q_k \bar D_l
\; ,  \\ 
\label{wrpv}
\eeq
along with a soft supersymmetry breaking and RPV term
\beq
{\cal L}^{\rm soft}_{\rm LV} \ = \ A_{j22} \tilde \ell_j \tilde \ell_2 \tilde
\mu_R^+ + {\rm (H.c.)}\; , \label{lsoft}
\eeq
where $j,k,l \in \{1,\ 3\}$ are the family indices, $Q_k$ and $L_k$ are
$k^{th}$ generation quark and lepton doublet superfields respectively, and
$D_k$ is the $k^{th}$ down-type quark singlet superfield. 
For the components of $\tilde \ell_j = (\tilde \nu_j, \tilde \ell^\pm_j)$,
there are two relevant $\lambda^\prime$-type vertices from Eq.~\eqref{wrpv}
which appear in Fig.~\ref{fig:main_diag},
viz. $\lambda'_{j11}(\tilde{\nu}_{j}d_{L}\bar{d}_{R}-\tilde{\ell}^\pm_{j}u_{L}\bar{d}_{R})$. Here
we have only considered first-generation quarks in the initial states, as
they have much larger parton distribution functions (PDFs) inside the proton
than 
the other two generations. We could replace all
of the family indices that are set to 2 in Eqs.~\eqref{wrpv} and \eqref{lsoft}
to  
a common but different value (as long as it were different to $j$), in which
case we would really have di-stau or di-selectron production (throughout this
paper, we implicitly include other
modes involving the accompanying sneutrinos). 
For definiteness, and because it can potentially explain the long-standing discrepancy of the SM prediction  with the measurement of the anomalous
magnetic moment of the muon (see e.g.~\cite{Miller:2012opa}),
we focus here on the di-smuon case, but bear in mind that the
other cases would lead to an identical signature at the LHC. 
The choice of $k$ and $l$ are irrelevant to the
gross phenomenology, since any choice results in light sub-jets. 
The choice of $j$ does affect whether one can obtain constraints and signals
from neutrinoless 
beta decay ($0\nu\beta\beta$)~\cite{Mohapatra:1986su, Hirsch:1995ek}. We shall
comment on this possibility later. We ban baryon 
number violating terms from
the RPV model (for example by using baryon
triality~\cite{Ibanez:1991hv}) in order to keep the proton stable, in  
accordance with observed lower limits on the proton decay
lifetime~\cite{Nishino:2009aa}. 
Other lepton number violating terms may be present, but should be sub-dominant
to the terms that we have written in Eqs.~\eqref{wrpv} and \eqref{lsoft} in
order 
for our analysis to be valid. 
We shall also set other sparticles not involved to be sufficiently heavy so
that they do 
not interfere with our analysis or the di-boson signal.

\begin{figure}[b!]
	\centering 
    \includegraphics[width=0.45\textwidth]{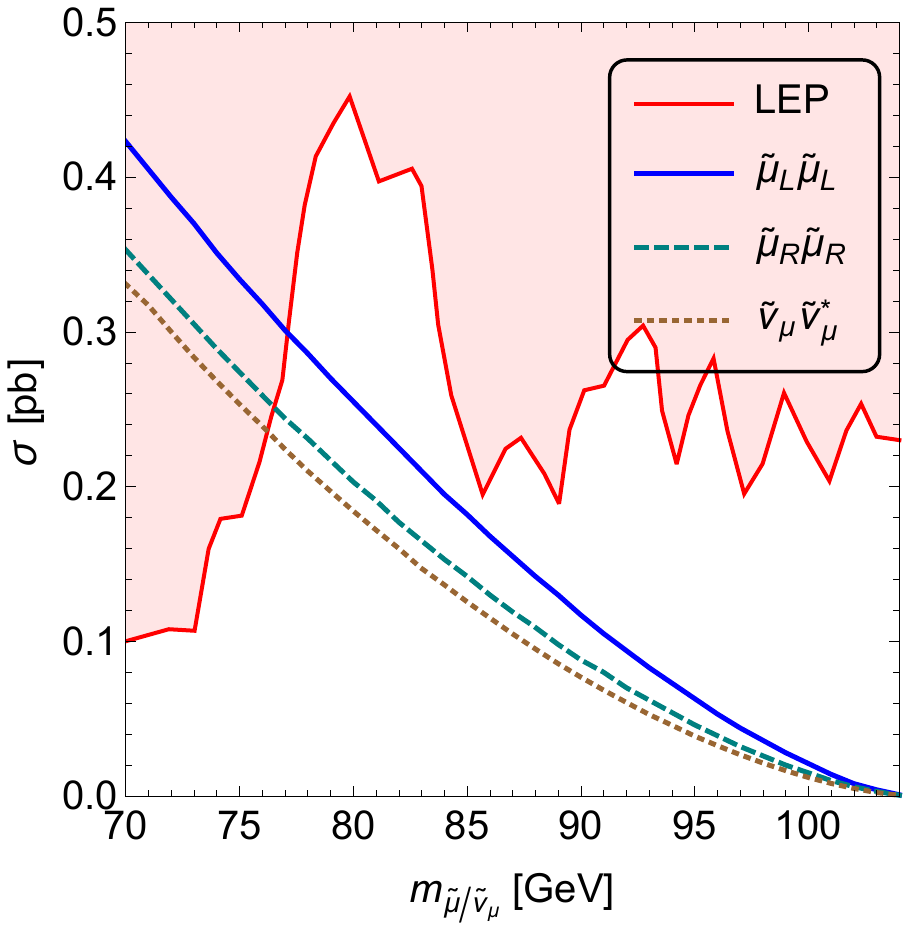}
    \caption{95$\%$ CL exclusion region (red shaded) derived from LEP II data for di-smuon production cross sections, followed by smuon decay into di-jets, as a function of the smuon mass. 
       For comparison, we also show the pair production cross-sections
      for $\tilde \mu_L$ (blue solid), $\tilde \mu_R$ (green dashed) and $\tilde \nu_\mu$ (brown dotted)
      at $\sqrt{s} = 209$ GeV. 
\label{fig:lims}}
\end{figure}
 The ATLAS di-boson analysis~\cite{Aad:2015owa} tagged a fat jet as a $W$ if its mass was in
the range $69.4<m_j/\textrm{GeV}<95.4$ after grooming and filtering, whereas
it was tagged as a $Z$ if $79.8 < m_j/\textrm{GeV} < 104.8$. There is clearly
an overlap between the $W$ and $Z$ tags, and therefore, the $WW$, $WZ$ and
$ZZ$ tagged regions are not completely disjoint (see Ref.~\cite{Allanach:2015hba} for a
detailed statistical analysis including the overlaps). We must make sure that
the smuons or muon sneutrinos in Fig.~\ref{fig:main_diag} are in the mass range
$69.4<m_{\tilde \mu}/\textrm{GeV}<104.8$ so that they are tagged as $W$
and/or $Z$ 
bosons. 
On the other hand, LEP II 4-jet searches~\cite{Abbiendi:2003rn} provide a lower bound on the smuon mass of around 77 GeV, as shown in Fig.~2, where we plot the 95\% CL exclusion limits for the 
di-smuon production cross-section at $\sqrt{s}=209$ GeV, as well as the corresponding model predictions, as a function of
the smuon mass assuming the smuons predominantly decay into di-jets.
The exclusion region depicted is on left-handed
smuons and is  more stringent than the one on right-handed smuons. The limit on
muon sneutrinos is, to a very good approximation, identical to the limit on
left-handed smuons. 
The corresponding cross section limits from $\sqrt
s=8$ TeV LHC (which are similar to the stop pair production limits) are even
weaker; see e.g.~\cite{Franceschini:2015pka}. 
% One might worry about single
% light sneutrino production at LEP and its subsequent decay to di-jets. However,
% for an 
% ${\cal 
%   O}$(80-90 GeV) sneutrino as in our case, the SM background is overwhelming and
% it is extremely  difficult to extract limits on RPV couplings using this
% signal~\cite{Dreiner:2012np}. 
%The lines display a prediction for an example scenario (see section~\ref{sec:ass}) for
%the LEP II production cross section of smuons and tau sneutrinos that we shall use to explain the ATLAS
%diboson excess. 
From Fig.~\ref{fig:lims} we see that if we have smuons with mass larger than 80
GeV, then our scenario should not fall afoul of the LEP II limits. 
We shall therefore
only use smuons in the range $80 < m_{\tilde \mu}/\textrm{GeV}< 105$. 

This possibility is intriguing because the necessarily light
smuons and muon sneutrinos will contribute to the anomalous magnetic moment of
the muon $(g-2)_\mu$, measurements of which~\cite{Bennett:2006fi} have long
been known 
to be discrepant with the SM at  
the 3.6$\sigma$ level, with $\delta a_\mu \equiv \delta (g-2)_\mu/2 = (2.9 \pm
0.8) \times 10^{-9}$~\cite{Miller:2012opa}.
However, SUSY gives one-loop contributions with smuons and neutralinos running
in a loop, along with a loop containing muon sneutrinos and charginos, yielding~\cite{Czarnecki:2001pv}
\beq
%\delta a_\mu \approx 1.3 \times 10^{-9} \left( \frac{100 \textrm{~GeV}
%}{ \widetilde m} \right)^2 \tan \beta,
\delta a_\mu \approx 1.3 \times 10^{-9} \left( \frac{100 \textrm{~GeV}
}{\textrm{min}(M_{\chi_1^\pm},\ M_{\chi_1^0})}\right)^2 \tan \beta \: {\rm sign}(\mu M_2),
\label{gm2}
\eeq
where the masses of smuons and muon sneutrinos are assumed to be around 100 GeV,   $M_{\chi_1^\pm}$ and $M_{\chi_1^0}$ are the masses of the lightest
chargino and the lightest neutralino, respectively, $\mu$ and $M_2$ are the Higgsino and wino mass parameters, respectively, and $\tan\beta\equiv v_u/v_d$ is the ratio of the up and down-type Higgs vacuum expectation values.\footnote{Using {\tt GM2Calc}~\cite{Athron:2015rva}, we have numerically verified that the linear dependence on $\tan\beta$ in Eq.~\eqref{gm2} is an approximation good to around 20\% for small to moderate $\tan\beta$ values, which will be the case for our benchmark point discussed later. For large $\tan\beta$, higher-order terms can become important and change this linear dependence~\cite{Marchetti:2008hw}.} 
%where $\widetilde m \equiv \min\left(\max(m_{\tilde \mu}, M_{\tilde
%    \chi_1^\pm}),\ \max(m_{\tilde \mu}, M_{\tilde \chi_1^0})\right)$ picks out
%the approximate dominant contribution~\cite{Czarnecki:2001pv}. 
Given that $\tan \beta 
\lsim 50$ from perturbativity and precision electroweak constraints~\cite{Cho:1999km}, whereas  $M_{\chi_1^\pm} \gtrsim 104$ GeV from LEP
constraints~\cite{Abbiendi:2003sc}, it appears 
that there is still plenty of viable parameter space where the discrepant
$(g-2)_\mu$ measurement is explained by sparticle loops.

One may worry that such light smuons would be ruled out by di-jet constraints
from the LHC \cite{Aad:2014aqa, Khachatryan:2015sja, Khachatryan:2015dcf, ATLAS:2015nsi},
through $q' \bar q \rightarrow \tilde{\mu} \rightarrow q' \bar q$, but in fact
the 
RPV coupling mediating such a process, $\lambda'_{2kl}$, may
be made small enough ($\lesssim 10^{-2}$) to relax the di-jet constraint through suppression of the
production cross-section, whereas the smuons in Fig.~\ref{fig:main_diag} will
always still decay into $q \bar q$ as long as there are no other competing
decay modes.  LHC di-jet constraints from resonant smuon
production~\cite{Dreiner:2012np} are proportional to $|\lambda_{211}'|^2$,
which we may set to be arbitrarily small without affecting the di-boson signal. 
The smuon width is
\beq
\Gamma(\tilde \mu \rightarrow \bar u d) \ = \  \frac{3}{16 \pi}
|\lambda'_{211}|^2 m_{\tilde \mu} \; ,
\label{eq:dstau}
\eeq
resulting in a decay length of 
\beq
{\rm L}/\textrm{cm}\ = \ 10^{-14} ( \bm{\beta \gamma} ) \left( \frac{100 \textrm{~GeV}}{m_{\tilde \mu}}
\right) \frac{1}{3 |\lambda'_{211}|^2}\; ,
\eeq
where $\bm{\beta}$ and $\bm{\gamma}$ are the usual relativistic kinematic factors. 
As long as $\lambda'_{211}>\mathcal{O}(10^{-6})$, the majority of the
decays should occur promptly. 

Another potential problem is the fact that  smuons and muon sneutrinos that
are too light
might be ruled out by precision electroweak constraints. Determining their
contributions to the electroweak parameters $S$ and
$T$~\cite{PhysRevLett.65.964}, Fig.~3 of 
Ref.~\cite{Cho:1999km} 
shows that even if {\em all three} left-handed slepton doublets have a mass of
100 GeV, at $\tan \beta=2$, one obtains $\Delta S=-0.05$ and $\Delta
T=0.0$, well within the 90$\%$ CL bound. With only one slepton
doublet required to be so light, we stay on the allowed side of the bound. 
In order to estimate whether the Tevatron may have ruled the scenario out,
we estimated the total production cross-section for smuons and muon sneutrinos
at $\sqrt s=1.96$ TeV proton-anti-proton collisions with {\tt
  Herwig7.0}~\cite{Bahr:2008pv,Bellm:2015jjp} to be 41 fb. Such a low
cross-section for a 4-jet final state is not excluded by any Tevatron search,
to the best of our knowledge.

%%%%%%%%%%%%%%
\section{The smuon masses and mixing}\label{sec:mass}

Without light right-handed neutrinos, the mass eigen-state coincides with the gauge eigen-state in the sneutrino sector.
The mass of the muon sneutrino is given by
\beq
m^2_{\tilde \nu_\mu} \ = \ m^2_{\tilde \ell_2} + \frac{1}{2}m_Z^2 \cos{2 \beta} \; ,
\eeq 
where $\tan\beta=v_u/v_d$ denotes the ratio of the vacuum expectation values of the two Higgs doublets $H_u$ and $H_d$ in the MSSM. 
In the gauge eigen-basis ($\tilde \mu_L, \tilde \mu_R$), the smuon mass matrix is given by
\beeq
M^2_{\tilde \mu} &\  = \ &   
\bmat
\tilde m^2_L & X_\mu \\
X_\mu & \tilde m^2_R
\emat \\ 
{\rm with} \quad \tilde m_L^2 & \ = \ & m_{\tilde \ell_2}^2 + m^2_\mu + m_Z^2\cos{2 \beta}\left(-\frac{1}{2} + \sin^2{\theta_w}\right)  \; , \nonumber \\
\tilde m_R^2 & \ = \ & m_{\tilde \mu_R}^2 + m^2_\mu - m_Z^2 \cos{2\beta}\sin^2{\theta_w} \; , \nonumber \\
X_\mu &\ = \ & m_\mu (A_\mu  - \mu \tan\beta)\; , \label{massTerms}
\eeeq
$\theta_w$ being the weak mixing angle.  
This mass matrix is diagonalized in the mass eigen-basis ($\tilde \mu_1, \tilde \mu_2$):
\beq
\bmat
m^2_{\smu2} & 0\\
0 & m^2_{\smu1} 
\emat
\ = \ U M_{\tilde \mu}^2 U^\dagger,\qquad 
U = \bmat c_{\thmu} &  s_{\thmu} \\ -s_{\thmu} & c_{\thmu} \emat,
\label{eq:stm}
\eeq
where $c_{\thmu}\equiv \cos\thmu$, $s_{\thmu} \equiv \sin \thmu$ and  
\beeq
&& m^2_{\tilde \mu_{2,1}}    =     
\frac{1}{2} \Big[ (\tilde m^2_L + \tilde m^2_R) 
\pm \sqrt{(\tilde m^2_L - \tilde m^2_R)^2 + 4 X_\mu^2 }
\Big], \label{eq:mass-eigen} \\
&& \tan 2 \thmu   =   \frac{2 X_\mu}{\tilde m^2_L - \tilde m^2_R} \; . 
\label{eq:tan2th}
\eeeq
%\beq
%\sin 2 \thtau = \frac{2 X_\tau}{m^2_{\stau2} - m^2_{\stau1}}~.
%\label{eq:sin2th}
%\eeq
%Alternatively one can also write

%\section{The Feynman rule}

The gauge and mass eigenbases are related by [cf. Eq.~\eqref{eq:stm}]
$\mu_L = c_{\thmu} \smu2 - s_{\thmu} \smu1$
and
$\mu_R = s_{\thmu} \smu2 + c_{\thmu} \smu1$.
Thus, one can find the Feynman rules for the vertices induced by Eq.~\eqref{lsoft} in the mass eigenbasis as
\beeq
\tilde \ell^-_j \tilde \nu_\mu \tilde \mu_R^+
&\ = \ & \tilde \ell_j^- \tilde \nu_\mu (s_{\thmu} \smu2^+ + c_{\thmu} \smu1^+) \; ,
 \\
\tilde \nu_j \tilde \mu_L^- \tilde \mu_R^+
% &=& A_{133} \tilde \nu_e  (\cos \thtau \stau2^- - \sin \thtau \stau1^-) (\sin \thtau \stau2^+ + \cos \thtau \stau1^+) 
% \nonumber \\
& \ = \ & \tilde \nu_j  (
  c_{\thmu} s_{\thmu} \smu2^+ \smu2^- 
- c_{\thmu} s_{\thmu} \smu1^+ \smu1^- \nonumber \\ &&
 \quad + \: c^2_{\thmu} \smu1^+ \smu2^- 
- s^2_{\thmu} \smu2^+ \smu1^-). 
\eeeq

Ignoring for now the di-jet decay mode via the $\lambda'_{j11} L_j Q_1
\bar{D}_1$ operator,  
the ratio of the partial decay widths of the slepton resonance to di-smuons can
be written as 
\beeq
 \tilde \ell^\pm_j  \ \to \ \tilde \nu_\mu \smu2^\pm  :   \tilde \nu_\mu \smu1^\pm
&  \ = \ & s^2_{\thmu} : c^2_{\thmu} \; ,
\label{eq:sele-decay}
\\
 \tilde \nu_j \  \to \ \smu2^+ \smu2^- :   \smu1^+ \smu1^-  :   \smu2^\pm \smu1^\mp
& \ = \ & \frac{s^2_{2 \thmu}}{4}   :   \frac{s^2_{2 \thmu}}{4}   :  1 - \frac{s^2_{2 \thmu}}{2} \; , \nonumber \\
\label{eq:snu-decay}
\eeeq
independent of the $A$-parameter in Eq.~\eqref{lsoft}. However, in practice, the $\lambda'$-terms in Eq.~\eqref{wrpv} also induce di-jet modes $\tilde \ell^\pm_j \to q\bar{q}'$ and $\tilde \nu_j \to q\bar{q}$, which cannot be neglected, since the same coupling is responsible for the production of the slepton resonance in Fig.~\ref{fig:main_diag}, and hence, cannot be arbitrarily suppressed. Thus, the relative branching ratios of $\tilde \ell_j$ to di-smus will depend on both $A_{j22}$ and $\lambda_{j11}$, as we will see in Section~\ref{sec:decay}.

\section{Mass assignment} \label{sec:ass}

In order to explain the di-boson excess by the $\tilde \ell_j$ production, two possibilities can be considered.
One is to make $\tilde \mu_2$ heavy and explain the di-boson excess by identifying fat $W/Z$ jets as fat $\smu1/\tilde \nu_\mu$ jets.  
If one requires the $\tilde \ell^\pm_j$ production contributes to the di-boson excess, $m_{\tilde \nu_\mu} \simeq \tilde m_L $ has to be around the gauge boson mass, $m_V \simeq (m_Z + m_W)/2$.
To remove the $\tilde \mu_2$ contribution from the signal, we need $\tilde m_R \gg m_V$.
Assuming $X_\mu \ll \tilde m_R$, the lighter smuon mass can be written as [cf. Eq.~\eqref{eq:mass-eigen}]
\beq
m^2_{\smu1} \ \simeq \ \tilde m_L^2 - \frac{2 X_\mu^2}{\tilde m_R^2 - \tilde m_L^2} \; .
\eeq
Therefore, to make $m^2_{\smu1}$ positive and also around the gauge boson mass, 
$X_\mu \ll \tilde m_R$ is indeed necessary.
From Eq.~\eqref{eq:tan2th} and knowing $\smu1 \sim \tilde \mu_L$,
one can see that $\theta_\mu \sim \frac{\pi}{2}$ and the couplings for the $\tilde \ell^\pm_j \to \tilde \nu_\mu \smu1^\pm$ and
$\tilde \nu_j \to \smu1^+ \smu1^-$ are suppressed by
$\cos^2 \thmu$ and $\sin^2 2 \thmu /4$, respectively. Thus in this case, 
the di-jet final state can more easily dominate the $\tilde \ell_j$ decay, instead of the di-smuon final state, disfavoring our RPV interpretation.

Another possibility is to bring down the masses of all particles in the smuon sector, i.e. $\tilde \nu_\mu, \smu1$ and $\smu2$, to around the average gauge boson mass scale $m_V$. If one demands $(m_{\smu2}, m_{\smu1}) \simeq (m_Z, m_W)$, both $\tilde m^2_L$ and $\tilde m^2_R$ have to be around the gauge boson mass scale, and 
$(\tilde m^2_L - \tilde m^2_R)$, $X_\mu$ have to be smaller than $m_V$, being related by
\beq
(m_Z^2 - m_W^2)^2 \simeq (\tilde m^2_L - \tilde m^2_R)^2 + 4 X^2_\mu \; .
\eeq
One can also find
\beq
s_{2 \theta_\mu} \ \simeq \ \frac{A_\mu  - \mu \tan \beta}{ 8651\,{\rm GeV} } \; ,\quad 
c_{2 \theta_\mu} \ \simeq \ \frac{ \tilde m^2_L - \tilde m^2_R }{ 1881 \,{\rm GeV^2} } \; .
\eeq
In this case, all decay modes in Eqs.~\eqref{eq:sele-decay} and \eqref{eq:snu-decay} 
are possible and the mixing can be suppressed as long as the smuon decays are prompt
because the $\tilde \mu_{1,2} \to q \bar q$ decay widths are proportional to 
$|\lambda'_{211} s_{\thmu}|^2$ and $|\lambda'_{211} c_{\thmu}|^2$, respectively.

For example, by taking
$\tan\beta = 1.5$, $m_{\tilde \ell_2} = 88$\,GeV, $m_{\tilde \mu_R}=80$\,GeV, 
$X_{\mu} = 537$\,GeV$^2$, we find
\beeq
m_{\tilde \nu_\mu} & \ = \ & 78.43\,{\rm GeV} \; , \nonumber \\
m_{\smu1} & \ = \ & 83.43\,{\rm GeV}\; ,  \nonumber \\
m_{\smu2} & \ = \ & 93.68\,{\rm GeV} \; , \label{eq:bench} \\
\sin \theta_\mu & \ = \ & 0.31 \; , \nonumber 
 \\
\tilde \ell_j^\pm \to \tilde \nu_\mu \smu2^\pm: \tilde \nu_\mu \smu1^\pm
& \ = \ & 0.097 : 0.903 \; ,
\nonumber \\
\tilde{\nu}_j \to \smu2^+ \smu2^- : \smu1^+ \smu1^- : \smu2^\pm \smu1^\mp
& \ = \ & 0.0874 : 0.0874 : 0.8252\; , \nonumber
\eeeq
We approximate $m_{\smu1}$ and
$m_{{\tilde\nu}_j}$ by $M_W$ and $m_{\smu2}$ by $M_Z$, since the values are
rather close. 
Eq.~\eqref{eq:bench} implies that $\tilde{\ell}_j^\pm$ mostly decays to $\tilde
\nu_\mu 
\smu1^\pm$ and $\tilde{\nu}_j$ mostly to $\smu2^\pm \smu1^\mp$, thereby
mimicking the $WW$ and $WZ$ final states, respectively, in the context of the
ATLAS di-boson search. 
We note here that by playing with the mass parameters, we could change the
effective proportions of $WW$, $WZ$ or indeed $ZZ$ final states. We shall here
stick to the approximation that $\tilde{\ell}_j^\pm$ always decays to 
$\tilde \nu_\mu \smu1^\pm$ and $\tilde{\nu}_j$ always decays to $\smu2^\pm
\smu1^\mp$. %\col{Also note that we use a low $\tan\beta$ for the benchmark spectrum quoted above, since this is preferred from electroweak precision constraints~\cite{Cho:1999km}, but high $\tan\beta$ up to about 50 is still compatible, as long as some $\tilde{\ell}_L$ are heavy.}

\section{Slepton decays} \label{sec:decay}
In order to explain the di-boson excess through the resonant slepton
production process of Fig.~\ref{fig:main_diag}, we assume $M = m_{\tilde
  \ell^\pm_j} \simeq m_{\tilde \nu_j} \simeq 1.9$\,TeV.\footnote{1.9 TeV provides a
good fit to the ATLAS di-boson excess and other CMS
data~\cite{Brehmer:2015cia, Dias:2015mhm}, although we could also have 
chosen 2 TeV, without affecting our main conclusions.} According to data, the resonance should not be too much wider than 100 GeV~\cite{Aad:2015owa} (although perhaps up to 160 GeV or so is still  acceptable).
The decay width of $\tilde \ell_j^\pm / \tilde \nu_j$ to the smuon and muon sneutrinos induced by $A_{2jj}$ from Eq.~\eqref{lsoft} is given by
\beeq
&& \Gamma( \tilde \nu_j \to \smu2^+ \smu2^-, \smu1^+ \smu1^-, \smu2^\pm \smu1^\mp )
  \ \simeq \  
\Gamma( \tilde \ell^\pm_j \to \tilde \nu_\mu \smu2^\pm, \tilde \nu_\mu \smu1^\pm) \nonumber \\ 
&& \qquad \  = \  \frac{|A_{j22}|^2}{16 \pi M} \left(1-\frac{4m^2_{\tilde{\mu}}}{M^2}\right)^{1/2}.
\label{eq:18}
\eeeq 
On the other hand, the decay width for the $q \bar q$ mode through the $\lambda'$-term in Eq.~\eqref{wrpv} is given by
\beq
\Gamma(\tilde \nu_j \to q \bar q) \ \simeq \ \Gamma(\tilde \ell^\pm_j \to q \bar q) 
 \ = \ 
\frac{3 |\lambda'_{j11}|^2 M}{16 \pi}.
\label{eq:19}
\eeq
The branching ratio for the $\tilde \nu_j$ to decay to smuons is therefore
given by 
\beq
{\rm BR}_{\tilde \mu} \  \simeq \ \frac{|A_{j22}|^2}{3 |\lambda'_{j11}|^2 M^2 +  |A_{j22}|^2} \; .
\label{eq:20}
\eeq
Here we have neglected the ${\cal O}(m^2_{\tilde \mu} / M^2 )$ terms, keeping
in mind the benchmark scenario given by Eq.~\eqref{eq:bench}.  
\begin{figure}
    \centering 
    \includegraphics[width=0.45\textwidth]{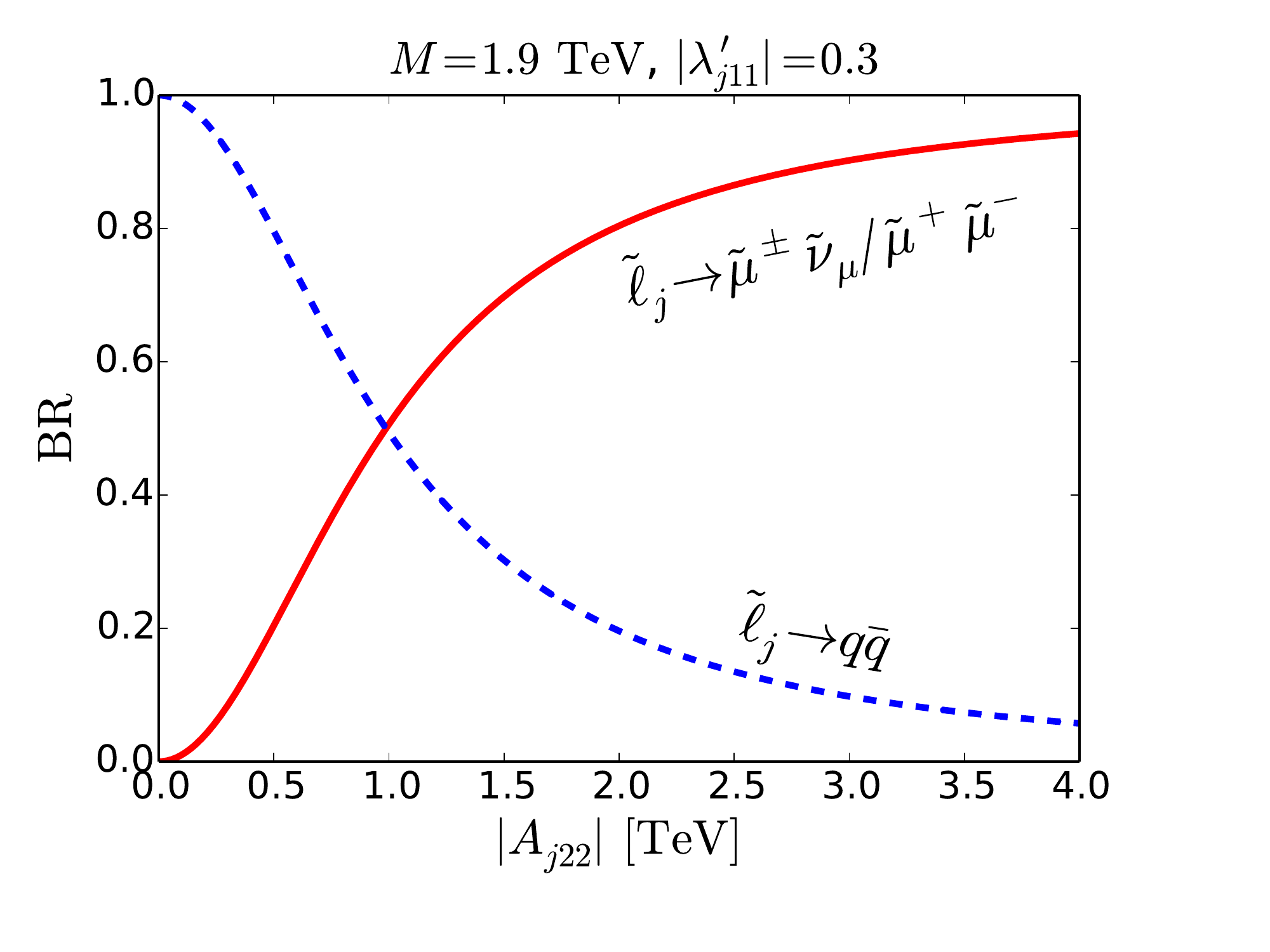}
    \caption{ Branching ratios of slepton decays to di-quark and di-smuon through RPV couplings given by Eqs.~\eqref{wrpv}
      and \eqref{lsoft}, respectively.    
\label{fig:BR}}
\end{figure}
Fig.~\ref{fig:BR} shows the branching ratios for
the smuon and di-jet final states as functions of $|A_{j22}|$,
where $M=1.9$\,TeV and $|\lambda'_{j11}|=0.3$ are assumed. It is clear that
for large $|A_{j22}|$, the di-smuon decay mode will dominate
over the di-jet mode, which is favorable for the di-boson excess.  We note
here that in principle, a very large $A_{j22}$-term could make the
soft-mass squared $(m_{\tilde \ell}^2)_{22}$ negative when running to higher
renormalization scales, since the renormalization group (RG) 
evolution of the slepton mass Lagrangian term $(m_{\tilde \ell}^2)_{22}$ has a
piece 
$d(m_{\tilde \ell}^2)_{22}/d \ln \mu =
-2|A_{j22}|^2/(16 \pi^2)+\ldots$~\cite{Allanach:2003eb}.
It is not clear immediately whether this
would 
destabilize the electroweak vacuum, since loop corrections to the energy
density of the minimum could be the same size as the tree-level
potential~\cite{Ellis:2008mc}. In order to calculate any such constraint
reliably, one can
use the ``RG-improved'' potential, taking the renormalization scale to be the
putative vacuum expectation value~\cite{Sher:1988mj}, which should in this
case be around 
the TeV scale (because it is driven by the TeV scale parameter
$A_{j22}$). Since this scale is not very far from the electroweak scale, there
one does not have to run the RG equations very far and the
constraints are likely to be weak because there is not much room for a
tachyonic smuon to arise. There are other potential directions in scalar field
space to check that are {\em not}\ associated with tachyons, but to reliably
calculate 
bounds from those, one would have to   
upgrade a package like {\tt Vevacious}~\cite{Camargo-Molina:2013qva} in order
to include RPV, which is beyond the scope of this paper and might be studied elsewhere.

We must also consider whether the model is non-perturbative, since, for
example, a loop correction to the quartic $\tilde l_j$ coupling may be large:
the dominant diagram is a box with smuons/muon sneutrinos running in the
loop. We find a one-loop correction to the coefficient of the effective potential term $V \supset  \lambda_{\tilde l_j}|\tilde{l}_j|^4$, where $\lambda_{\tilde l_j}=(g_2^2+g_1^2)/4$ at the tree-level ($g_2$ and $g_1$ being the $SU(2)_L$ and $U(1)_Y$ gauge couplings, respectively)~\cite{Okada:1990gg, Cho:2006sm}
\begin{equation}
\Delta \lambda_{\tilde l_j} \approx -\frac{1}{384 \pi^2}
\left(\frac{A_{j22}}{\tilde m}\right)^4, \label{eq:quar}
\end{equation}
assuming a common mass $\tilde m$ for the left-handed smuons and muon sneutrinos.
Below, we shall impose that this correction is not too large, i.e.
\begin{equation}
 |\lambda_{\tilde l_j}+\Delta
\lambda_{\tilde l_j}| \ < \ 4 \pi\; ; \label{eq:pert}
\end{equation} 
otherwise, the theory would be non-perturbative
and we would not be able to trust the accuracy of our results. 

One might also worry whether the negative sign in Eq.~\eqref{eq:quar} leads to a 
    charge-breaking minimum (CBM) in the direction of the slepton~\cite{Abel:1998ie, Dedes:2005ec}. For a robust determination of whether this is unsafe for us, we need to compute the lifetime of this minimum. Here we will simply use a conservative constraint by demanding that the coefficient of the quartic-term in the one-loop effective potential is positive-definite, i.e.  
\begin{equation}
\lambda_{\tilde l_j}+\Delta
\lambda_{\tilde l_j} \ \geq \ 0 \; . 
\label{eq:cbm}
\end{equation}

In any case, our RPV
scenario with a light smuon is consistent with all current experimental
constraints~\cite{Barbier:2004ez} and may be tested soon in the ongoing Run II
phase of the LHC. 
For a detailed discussion of light smuon phenomenology at the LHC, see
e.g.\ Refs.~\cite{Allanach:2003eb, Desch:2010gi}.

\section{Fitting the Di-boson Excess} \label{sec:fit}
\begin{figure}[t!]
    \centering 
    \includegraphics[width=0.45\textwidth]{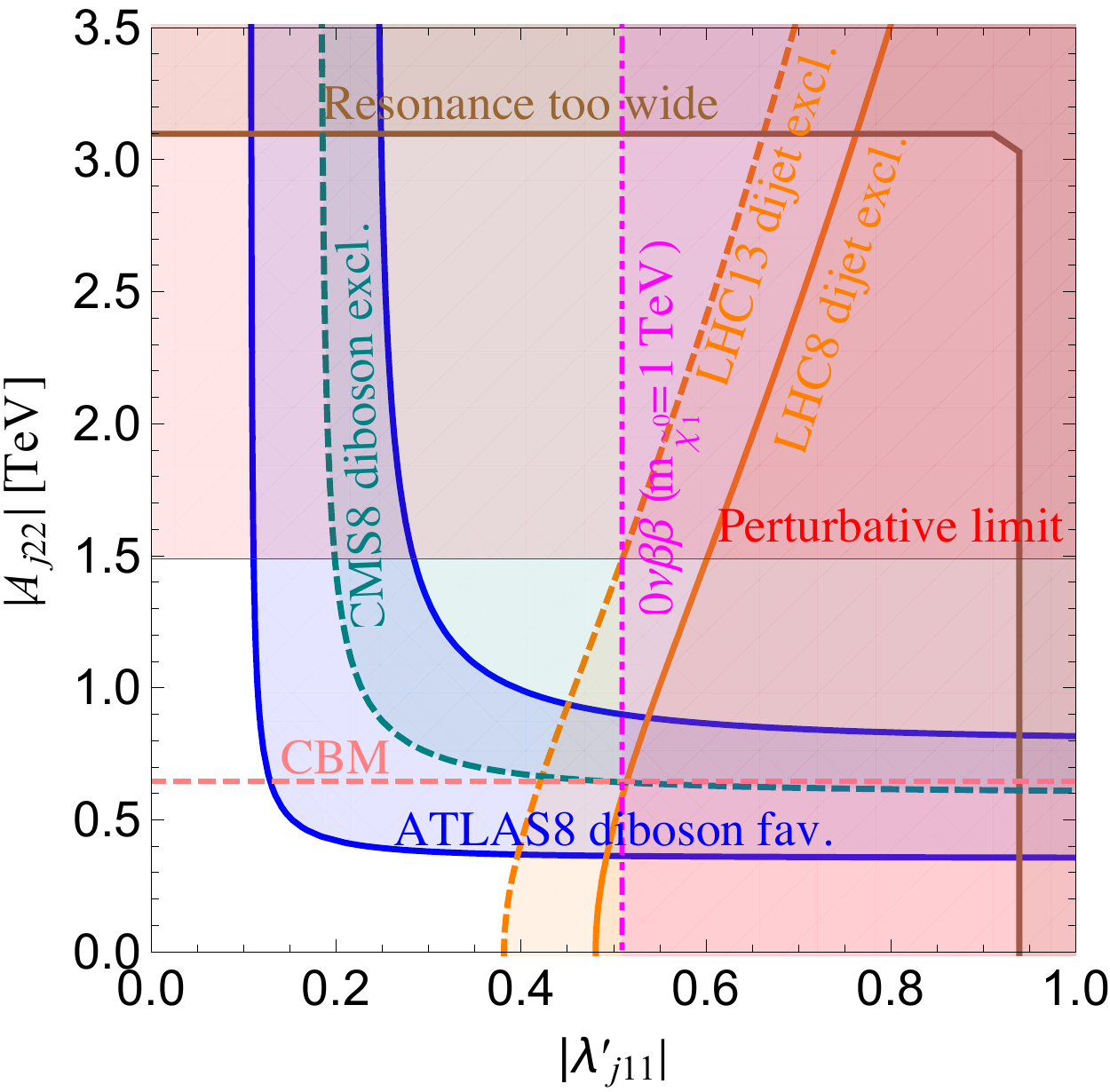}
    \caption{The RPV parameter space favored by the ATLAS di-boson excess in 8 TeV LHC data 
      (blue shaded region). The 8 TeV exclusion regions from CMS di-jet (solid orange) and di-boson (green), as well as the 13 TeV exclusion from CMS di-jet (dashed orange) searches are also shown. The magenta dashed
      vertical line shows the upper limit for the $j=1$ case on
      $|\lambda'_{111}|$ due to null results from a recent $0\nu\beta\beta$
      search, assuming the lightest neutralino mass of 1 TeV.
      The light red region in the top half of the plot is non-perturbative as
      estimated by Eq.~\eqref{eq:pert} and the region
      outside of the solid brown line has a resonance width larger than 100 GeV. The horizontal pink dashed line is the suggestive upper bound for $|A_{j22}|$ obtained from Eq.~\eqref{eq:cbm} beyond which one might develop a charge-breaking minimum. 
\label{fig:diboson}}
\end{figure}

%We interpret the ATLAS excess in the di-boson channel $pp\to X\to VV \to 2$ fat jets~\cite{ATLAS1} using the resonant di-stau process shown in Figure~\ref{fig:main_diag}.  For the benchmark scenario given in Eq.~\eqref{eq:bench}, we expect the $\tilde{e}_L\to \tilde{\tau}_1\tilde{\tau_2}$ decay to mimick the $WZ$ channel, whereas the $\tilde{\nu}_e\to \tilde{\nu}_\tau \tilde{\tau}_1$ to mimick the $WW$ channel. 
We first calculate the resonant production cross sections for a 1.9 TeV $\tilde \ell^\pm_j$ or $\tilde{\nu}_j$ at $\sqrt s=8$ TeV LHC using the RPV model implementation in {\tt FeynRules}~\cite{Alloul:2013bka} and the parton-level event generation in {\tt MadGraph5}~\cite{Alwall:2014hca} with {\tt NNPDF2.3} leading order PDF sets~\cite{Ball:2012cx}.  We get 
\begin{align}
\sigma(pp\to \tilde{\ell}_j^\pm) \ = \ 75~{\rm fb} \; ,\quad \sigma(pp\to \tilde{\nu}_j, \tilde{\nu}_j^*) \ = \  359~{\rm fb} \; , \nonumber 
\end{align}
normalized to $|\lambda'_{j11}|^2=1$. 
The decay width of $\tilde{\ell}_j \to \tilde{\nu}_\mu^\pm \tilde{\mu}, ~ \tilde \mu^+ \tilde \mu^-$ is given by Eq.~\eqref{eq:18} and that of $\tilde{\ell}_j \to q \bar q$ is given by Eq.~\eqref{eq:19}. 
These are assumed to be the dominant decay modes so that the branching ratio
to di-smuon is given by Eq.~\eqref{eq:20}. The smuons (and muon sneutrinos) are
assumed to decay into di-jets with a 100\% branching ratio, as argued below
Eq.~\eqref{eq:dstau}, which is reasonable given that these are the lightest
sparticles in the model. 
Ref.~\cite{Allanach:2015hba} unfolded cross-contamination of the $WW$, $WZ$
and $ZZ$ channels and estimates of the efficiencies to bound the case where 
one has contributions from all three channels. Here, 
after the
approximations listed under Eq.~\eqref{eq:bench}, we have contributions to the
$WW$-like channel from charged slepton production and from the $WZ$-like
channel from sneutrino production, whereas we neglect any $ZZ$-like
channel production. By referring to the right-hand panel of Fig.~4 of
Ref.~\cite{Allanach:2015hba}, 
we see that the ATLAS constraint on the sum of 
$WW+WZ$ channels production cross section times branching ratio should be  
approximately $5-25$ fb to 95$\%$ CL.
Our prediction for this quantity is
\beq
 \sigma_{\rm sig.} \ = \ |\lambda'_{j11}|^2 \: {\rm BR}_{\tilde{\mu}} \big[\sigma(pp\to \tilde \ell^\pm_j)
+\sigma(pp\to \tilde{\nu}_j)\big] .
\label{eq:signal}
\eeq
This favored region (`ATLAS8 diboson fav.') is shown
by the blue shaded region in Figure~\ref{fig:diboson}. The corresponding CMS
search for boosted di-bosons~\cite{Khachatryan:2014hpa} has given a stringent
95\% CL upper limit of 14.3 fb on the signal cross section for 1.9 TeV
invariant mass, which excludes  the green shaded region in
Figure~\ref{fig:diboson}. Note that this is still consistent with a large part
of the  
parameter space favoring the ATLAS di-boson excess. Furthermore, there is
another stringent constraint coming from the LHC di-jet
searches performed in Run-I~\cite{Aad:2014aqa, Khachatryan:2015sja},  and more recently, in early Run-II~\cite{Khachatryan:2015dcf, ATLAS:2015nsi}, which are also applicable to
$pp\to \tilde \ell_j^\pm /\tilde \nu_j \to q \bar q$. At $\sqrt s=8$ (13) TeV LHC, the cross section for a  1.9 TeV $q \bar q$ resonance must be smaller than 100 (400) fb~\cite{Khachatryan:2015sja, Khachatryan:2015dcf},
which excludes at 95\% CL the solid (dashed) orange shaded region in
Figure~\ref{fig:diboson}. There are also theoretical constraints from perturbativity [cf.~Eq.~\eqref{eq:pert}] and CBM [cf.~Eq.~\eqref{eq:cbm}], which are shown in Figure~\ref{fig:diboson} by the horizontal red solid and pink dashed lines, respectively. We have not shaded the CBM region, since the CBM bound shown here should not be considered as a strict upper limit, unless and until one does a lifetime calculation, which is beyond the scope of this paper. In any case, we find that there still survives a resonable portion of the parameter space
in our RPV scenario consistent with the ATLAS di-boson
excess.  

We note here that for the $j=1$ case the $\lambda'_{111}$ coupling also induces  neutrinoless double beta decay ($0\nu\beta\beta$)~\cite{Mohapatra:1986su, Hirsch:1995ek}, and hence, constrained by the current experimental limits on $0\nu\beta\beta$ half-life. Using the latest 90\% CL combined limit for $^{76}$Ge isotope from GERDA phase-I~\cite{Agostini:2013mzu}, we find~\cite{Faessler:1996ph, Allanach:2009xx}
\begin{align}
|\lambda'_{111}| \ \lesssim \  4.5 \times 10^{-4}\left(\frac{m_{\tilde{e}_L}}{100~{\rm GeV}}\right)^2\left(\frac{m_{\tilde{\chi}_1^0}}{100~{\rm GeV}}\right)^{1/2} .
\label{eq:ndbd}
\end{align}
For a selectron mass of 1.9 TeV as required here to explain the ATLAS di-boson excess, we obtain a mild upper limit of $|\lambda'_{111}|\lesssim 0.51$ for the lightest neutralino mass $m_{\tilde{\chi}_1^0}=1$ TeV, as shown by the dashed vertical line in Figure~\ref{fig:diboson}. From Eq.~\eqref{eq:ndbd} and Fig.~\ref{fig:diboson}, we can readily infer that the $0\nu\beta\beta$ constraint still allows some parameter space favoring the ATLAS di-boson excess as long as the lightest neutralino is heavier than 150 GeV in our RPV scenario. 
If we were to re-fit for a slepton mass of 2 TeV, this line would be less
restrictive and move to the right. The CMS dijet bound would moves slightly
toward the left and there are other small changes, but the qualitative picture
as shown in the Figure remains. 
% We should also emphasize that the light stau scenario invoked here is
% consistent with the cross section limits for stau pair production from LEP
% 4-jet searches~\cite{Abbiendi:2003rn}. 

\section{Discussion and Conclusion} \label{sec:conc}
Before concluding, we wish to make a few comments on the testability of our
scenario and its applicability to other potentially relevant excesses with
respect to the SM predictions: \\

\noindent
(i)  The di-boson interpretation of the ATLAS excess inevitably leads to
leptonic and semi-leptonic final states, along with the hadronic decays of the
di-boson system. In contrast, our model, as currently written, does not
predict leptonic decays of the smuons, thus providing a potential explanation
of the absence of a corresponding di-boson excess in the leptonic or
semi-leptonic channels~\cite{Khachatryan:2014gha, ATLAS-CONF-2015-045}. With
more statistics pouring in from the Run II phase of the LHC, this will be a
clear distinguishing feature of our scenario in the near future. \\

\noindent
(ii) Unlike the $W'$ interpretation of the ATLAS di-boson excess which involves $WZ$ final states (see e.g.~\cite{Hisano:2015gna,Dobrescu:2015qna,Brehmer:2015cia,Allanach:2015hba,Dev:2015pga}), and hence, necessarily leads to a $WH$ excess ($H$ being the SM Higgs boson) by virtue of the Goldstone equivalence theorem, our
ATLAS di-boson favored region in Fig.~\ref{fig:diboson} does not suffer from
any such constraints. 
On the other hand, a recent CMS search~\cite{CMS-PAS-EXO-14-010} seems to
suggest a mild global excess of 1.9$\sigma$ in the $WH$ channel with $H \to b
\bar b$ and $W \to \ell \nu$.\footnote{The corresponding ATLAS
    search~\cite{Aad:2015yza} and a similar CMS search in the all-hadronic
    final states~\cite{Khachatryan:2015bma} have not reported any such excess
    though.}
If this excess becomes statistically significant, one way of accommodating it in our model is to have $m_{\tilde \nu_\mu} \simeq m_H$ and assume $\tilde \nu_\mu$ predominantly decays to $b \bar b$ through
the $\lambda^\prime_{233}$ coupling.
The leptonic $W$ decay can also be mimicked by 
augmenting Eq.~\eqref{wrpv} with another RPV term $\lambda_{2kl}{L_2 L_k \bar E_l}$, where $k,l \in \{1,\ 3\}$, which will induce a non-zero branching ratio of $\tilde \mu^\pm \to \ell^\pm_l \nu_k$.
%The Higgs decay can be easily micicked by the tau sneutrino, if its mass is close to the Higgs mass. As far as the leptonic W decay is concerned, as it is currently written, our model does not predict leptonic
%decays of the stau.
%However, one way to accommodate the WH excess is by augmenting
%Eq.~\eqref{wrpv} by another RPV coupling $W \supset \lambda_{3kl}{L_3 L_k \bar E_l}$ in order to obtain a non-zero branching ratio to leptons. 
%This could also lead to interesting multi-lepton signatures of light staus at the LHC~\cite{Desch:2010gi}. 
Unlike $W$ decays, we do not generically expect the leptonic decays of smuons to be flavor universal.
% Generically, we would not expect
% such leptonic decays to be flavor universal, nor would we expect them to match
% the leptonic branching fraction of $W$ bosons. 
Therefore, this could serve as
another distinguishing feature of our scenario. Yet another difference with respect to the $WZ$ final state is that it leads to a mono-jet signature in the decay channel $Z \to \nu \bar \nu$ and $W \to $a fat jet~\cite{Liew:2015osa}, whereas our scenario does not predict any such signatures.\\ 

\noindent
(iii) In the case where we choose the indices labeled as `2' in
Eqs.~\eqref{wrpv} and \eqref{lsoft} to instead be `3', we have di-stau (or tau
sneutrino) production. In
this 
case, the analysis of the collider phenomenology proceeds similarly to the
smuon case: the LEP constraints 
are identical. However, in the case of di-stau production, one does not
address the discrepancy between the measurement and the SM prediction of
$(g-2)_\mu$.\\

Interestingly, our scenario might explain some other Run I excesses, as follows: 
CMS searches for a right-handed charged gauge boson reported a
$2.8\sigma$ excess in the $eejj$ final state~\cite{Khachatryan:2014dka}. In
addition, the CMS search for di-leptoquark production has found a $2.4\sigma$
excess in the $eejj$ channel and a $2.6\sigma$ excess in the $e\nu jj$
channel~\cite{CMS:2014qpa}. It is possible to explain both of these excesses
with 
resonant slepton production in RPV SUSY, which decays to a lepton and a
chargino/neutralino, followed by three-body decays of the neutralino/chargino
via an RPV  coupling~\cite{Allanach:2014lca, Allanach:2014nna}.  In principle,
it is also 
possible to simulataneously accommodate the ATLAS di-boson excess in this
scenario, e.g.~ $pp\to \tilde{\ell}^\pm_1 \to
e\tilde{\chi}^0_1\to ee\tau \tilde{\tau} \to eejj + p_T^{\rm miss}$. %\slashed{E_T}$. 
However, there are three potential problems with this solution: (a) the
leptonic decay of the tau also gives muons, so we would expect $e \mu jj$ final
states as well, which does not show any significant excess at the LHC so far;
(b) the di-jets 
from a light smuon decay tend to be highly  boosted, as discussed above; so the
signal efficiency will drop drastically if we require well-separated jets to
explain the CMS excesses; (c) the CMS excess in $eejj$ favors more
opposite-sign di-electron final states, whereas the Majorana neutralino decays
produces same-sign electrons with the same rate as opposite sign. A detailed
analysis addressing 
these issues is beyond the scope of the current paper and is left for a
future study.

In conclusion, we have presented a new supersymmetric interpretation of the
ATLAS di-boson excess within an $R$-parity violating low-scale SUSY
framework that {\em ab~initio}\ does not predict leptonic branchings, although
the model can be tweaked to predict them. In fact, a recent combination by
ATLAS of its channels~\cite{Aad:2015ipg} shows that, once the leptonic
channels are 
added, the global significance of the di-boson excess (assuming that the
events consist of real $WW$, $ZZ$ or $WZ$) goes down, indicating a better fit
with only hadronic decays. 
In particular, we propose a
sparticle spectrum with smuon masses in 
the 80-105 GeV range and a resonant 2 TeV slepton that can be tested in the
Run II phase of the LHC. These necessarily light smuons are then in turn
linked to the discrepancy between the measurement and SM prediction of the
anomalous magnetic moment of the muon. 
If the di-boson excess persists and becomes statistically significant, it could
potentially be the first sign of SUSY at the LHC.

\section{Acknowledgments}
This work of B.C.A. has been partially supported by STFC grant
ST/L000385/1. The work of P.S.B.D. is supported in part by a TUM University
Foundation Fellowship and the DFG cluster of excellence ``Origin and Structure
of the Universe". B.C.A. and K.S.  
would like to thank TUM for hospitality extended during the conception of this
work. We thank Ben O'Leary for helpful discussions regarding electroweak
stability. 

\bibliographystyle{apsrev}
\bibliography{note}

\end{document}